\documentclass[prl,aps,showpacs,twocolumn]{revtex4}
\usepackage{amsfonts}
\usepackage{amsmath}
\usepackage{amssymb}
\usepackage{graphicx}%
\begin{document}

\title{Maximal Success Probabilities of Linear-Optical Quantum Gates}%

\author{Dmitry B. Uskov$^1$}%
\author{Lev Kaplan$^1$}%
\author{A. Matthew Smith$^1$}%
\author{Sean D. Huver$^2$}
\author{Jonathan P. Dowling$^2$}
\affiliation{$^1$Department of Physics, Tulane University, New Orleans, Louisiana 70118 \\
$^2$Hearne Institute for Theoretical Physics, Department of Physics and Astronomy, Louisiana State
University, Baton Rouge, Louisiana 70803}%

\date{\today}

\begin{abstract}

Numerical optimization is used to design linear-optical devices that implement a desired quantum gate with perfect fidelity, while maximizing the success rate. For the 2-qubit CS (or CNOT) gate, we provide numerical evidence that the maximum success rate is $S=2/27$ using two unentangled ancilla resources; interestingly, additional ancilla resources do not increase the success rate. For the 3-qubit Toffoli gate, we show that perfect fidelity is obtained with only three unentangled ancilla photons -- less than in any existing scheme -- with a maximum $S=0.00340$. This compares well with $S=(2/27)^2/2 \approx 0.00274$, obtainable by combining two CNOT gates and a passive quantum filter~\cite{Ralph2007}. The general optimization approach can easily be applied to other areas of interest, such as quantum error correction, cryptography, and metrology~\cite{wildeuskov, Durkin2007}.
\end{abstract}


\maketitle
Linear optics is considered as a viable method for scalable quantum information processing, due in large part to the seminal work of Knill, Laflamme, and Milburn (KLM)~\cite{KLM01}. These authors showed that an elementary quantum logic gate on qubits, encoded in photonic states, can be constructed using a combination of linear-optical elements and quantum measurement. The trade-off in this measurement-assisted scheme is that the gate is properly implemented only when the measurement yields a positive outcome, i.e., the gate is non-deterministic. Soon after the KLM scheme became a paradigm for linear optical quantum computation (LOQC), it became clear that there is a general unresolved theoretical problem of finding the optimal implementation for a desired quantum transformation~\cite{VanMeter06}.

For the nonlinear sign (NS) gate, which acts on photons in a single optical mode, $ \alpha_0 |0 \rangle +  \alpha_1 |1 \rangle + \alpha_2 |2 \rangle \to \ \alpha_0 |0 \rangle +  \alpha_1 |1 \rangle - \alpha_2 |2 \rangle $, the maximum success probability without feed-forward has been theoretically proved to be 1/4~\cite{EisertNS05}. Here we focus on more complicated gates, taking as examples the two-qubit controlled sign (CS) gate (equivalently, the CNOT gate), and the three-qubit Toffoli gate. For these physically important gates, existing theoretical results are limited to upper or lower bounds on the success probability~\cite{KnillCZ02, Knill03, Ralph2007}.

A linear-optical quantum gate, or state generator (LOQSG)~\cite{VanMeter06}, can be viewed formally as a device implementing a contraction transformation (for ideal detectors) that converts pure input states into desired pure output states. The goal of the optimization problem is to find a proper linear optical network (see Fig. 1), characterized by a unitary matrix $\mathbf{U}$, that performs the desired transformation~\cite{Reck94, Kok07}. The problem is naturally partitioned into two tasks: i) finding a subspace of perfect fidelity within the space of all unitary matrices $\mathbf{U}$, and ii) maximizing the success probability within this subspace. While in this paper we address transformations implemented by linear optics, the method is universal and with minor modifications can be successfully applied to any quantum-information problem involving unitary operations combined with measurements.

\begin{figure}[ht]
\includegraphics[width=8.4cm]{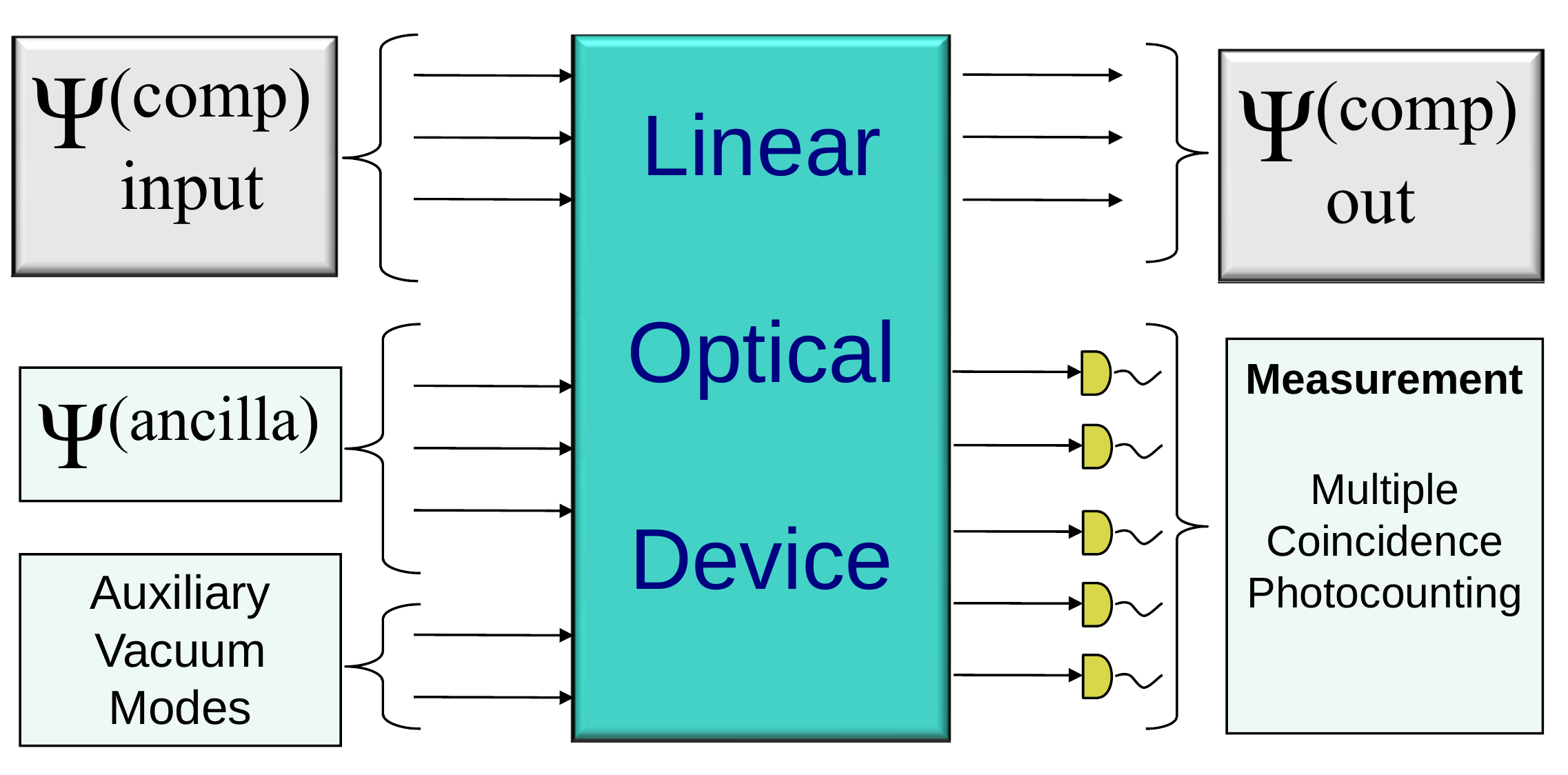}
\caption{\label{figdevice} A general measurement-assisted transformation (e.g., a quantum logic gate or a linear-optical quantum state generator).}
 \end{figure}

Originally, the linear-optical device was envisioned as a network of linear-optical elements~\cite{kwiat2008}, as for example in the original KLM scheme, where the CS gate is constructed as a combination of two NS gates~\cite{KLM01}. In practice, a functional microchip-based device may instead considered as an integrated light circuit~\cite{kwiat2008}, performing one large operation, as shown schematically in Fig.~\ref{figdevice}. Here the input state $|\Psi_{\rm in}\rangle =|\Psi_{\rm in}^{\rm comp}\rangle \otimes |\Psi_{\rm in}^{\rm ancilla}\rangle \otimes |\Psi^{\rm vacuum}\rangle$ is a product of a computational input state, an ancilla state, and possibly a vacuum state. Assuming dual-rail encoding, the computational state will consist of $M_c$ photons in $N_c=2M_c$ optical modes encoding an arbitrary state of $M_c$ qubits, e.g., the logical two-qubit state $|\uparrow\rangle \otimes |\uparrow\rangle$ may be represented in four optical modes by $|1_1,0_2 \rangle \otimes |1_3,0_4\rangle$ in the Fock basis. The ancilla input state of $M_a$ photons distributed over $N_a$ modes may be a separable state, an entangled state, or even half of an ebit state carrying spatially distributed entanglement, as required for example in entanglement-assisted error correction~\cite{Brun2006, wildeuskov}. Finally, $N_v$ auxiliary vacuum modes contain zero photons in the initial state.

The core of the device is the transformation $a_i^{({\rm in})\dag} \rightarrow U_{i,j} a_j^{({\rm out})\dag}$ of the photon creation operators between the input and output states. Here $\mathbf{U}$, which contains all physical properties of the device, is an $N \times N$ unitary matrix, where $N=N_c+N_a+N_v$ is the total number of modes. The matrix ${\mathbf{U}}$ associated with the physical device induces a transformation $\mathbf{\Omega}$ acting on the input state, where $\mathbf{\Omega}$ is a high-dimensional irreducible representation of ${\mathbf{U}}$~\cite{Perelomov86}.
Writing the total input state in the Fock representation as $|\Psi_{\rm in} \rangle = |n_{1}, n_2, \ldots, n_N \rangle$, where $\sum n_i=M_c+M_a$ is the total number of photons, $\mathbf{\Omega}$ takes the form

\begin{equation}
\label{Eq:OutState}
|\Psi_{\rm out}\rangle = \mathbf{\Omega(U)} |\Psi_{\rm in}\rangle = \prod_{i=1}^{N} \frac{1}{\sqrt{n_{i} !}} \left( \sum_{j=1} U_{i,j} a_{j}^{({\rm out})\dag} \right)^{n_i}  |0 \rangle \,.
\end{equation}

Next, a measurement is applied to the $N\!\!-\!\!N_c$ ancilla and vacuum modes. This measurement is formally described by a Kraus POVM operator acting on these modes only: $\mathbf{P} = |0_{N_c+1}, 0_{N_c+2},\ldots, 0_{N}\rangle \langle \Psi_{\rm measured}|$. In the most natural case of a photocounting measurement, $\langle \Psi_{\rm measured}|=\langle k_{N_c+1}, k_{N_c+2}, \ldots, k_{N}| $, where $k_i$ is the number of photons measured in the $i$-th mode. Finally, the resulting transformation of the computational state is a contraction quantum map  $|\Psi_{\rm out}^{\rm comp} \rangle = \mathbf{A} |\Psi_{\rm in}^{\rm comp} \rangle/\| \mathbf{A} |\Psi_{\rm in}^{\rm comp} \rangle\| $~\cite{Kraus83}, where $\mathbf A ( \mathbf U ) $ is defined by
\begin{equation}
\label{Eq:proj}
\mathbf{A(U)} | \Psi_{\rm in}^{\rm comp} \rangle = \langle k_{N_c+1}, k_{N_c+2}, \ldots, k_{N}| {\mathbf{\Omega(U)}} |\Psi_{\rm in}\rangle\,.
\end{equation}
The linear operator $\mathbf{A}$ contains all the information of relevance to the gate or state transformation.

Now we consider the main properties of Eq.~(\ref{Eq:proj}) relevant to the optimization problem. In the Fock basis, the matrix $\mathbf A$ is a submatrix of the larger matrix $\mathbf{\Omega}$, and in accordance with Eq.~(\ref{Eq:OutState}),  matrix elements of $\mathbf A$ are given as polynomials of degree $M_c+M_a$ in the matrix elements of $\mathbf{U}$.
For example, the matrix element for transforming the two-qubit computational state $|\uparrow\rangle \otimes |\uparrow\rangle$ into $|\uparrow\rangle \otimes |\downarrow\rangle$, with two ancilla photons in modes 5 and 6, is
\begin{equation}
\langle 1_1 0_2 0_3 1_4 |\mathbf A|1_1 0_2 1_3 0_4 \rangle=\!\!\!\!\!\!\!\!\!
\mathop{\sum_{j_1,j_2,j_3,j_4}}_{={\rm permutations} \; (1,4,5,6)} \!\!\!\!\!\!\!\!\!
U_{1,j_1} U_{3,j_2} U_{5,j_3} U_{6,j_4} \,.
\end{equation}
More generally, in the Fock representation, all matrix elements are calculated as permanents of $\mathbf{U}$~\cite{VanMeter06}.

Furthermore, if the total number of measured photons $\sum_{i={N_c+1}}^{N} k_i$ equals the number of input ancilla photons $\sum_{i={N_c+1}}^{N} n_i$, then Eq.~(\ref{Eq:proj}) leaves the number of computational photons invariant. The dual-rail computational basis is a subset of all possible states of $M_c$ photons in the $2M_c$ computational modes. Thus, the transformation matrix $\mathbf{A}$ is a rectangular matrix, mapping the computational Hilbert space, of dimension $2^{M_c}$, to a larger Hilbert space, of dimension $(3M_c-1)!/(2M_c-1)!(M_c)!$.

We now define precisely the operational fidelity of a transformation, which in general differs from the common measure of fidelity for a state transformation~\cite{Nielsen00}. Physically, the transformation $\mathbf{A}$ has $100 \% $  fidelity  if it is proportional to the target transformation operation $\mathbf{A}_{\rm t}$, i.e., $\mathbf{A} = g \mathbf{A}_{\rm t}$, where $g$ is an arbitrary complex number (in which case $S= |g|^2$ is the success probability of the transformation~\cite{Lapaire03}). In general, we may
consider complex rays $\beta_1\mathbf{A} $ and $\beta_2\mathbf{A}_{\rm t} $,  $\beta_1 , \beta_2 \subset \mathbb{C} $ as elements of a complex projective space, and
define the fidelity as
\begin{equation}
\label{Eq:F}
F(\mathbf U)=\frac{\langle \mathbf{A} | \mathbf{A}_{\rm t} \rangle \langle \mathbf{A}_{\rm t} | \mathbf{A} \rangle}{\langle \mathbf{A} | \mathbf{A} \rangle \langle \mathbf{A}_{\rm t} | \mathbf{A}_{\rm t}\rangle } \,,
\end{equation}
where $\mathbf{A} \equiv \mathbf A ( \mathbf U ) $  is defined by Eq.~(\ref{Eq:proj}). The Hermitian inner product is $\langle \mathbf A | \mathbf B \rangle \equiv {\rm Tr}(\mathbf A^{\dag} \mathbf B )/D_c$, and $D_c=2^{M_c}$ is the dimension of the computational space. $F$ is closely related to the
Fubini-Study distance $\gamma=\cos^{-1}(\sqrt{F})$~\cite{Bengtsson2006}, but for numerical computations $F$ has the advantage of being non-singular near $F=1$.

In general, the success probability $S$ depends on the initial state $|\Psi_{\rm in}^{\rm comp} \rangle$. $S$ is bounded above by the square of the operator norm, $\| \mathbf{A} \|^2 \equiv  \{ \mathbf{A} \}_{\rm max}^2 = {\rm Max}(\langle \Psi_{\rm in}^{\rm comp} | \mathbf{A}^{\dag} \mathbf{A} | \Psi_{\rm in}^{\rm comp} \rangle)$, and below by $\{ \mathbf{A} \}_{\rm min}^2 = {\rm Min}  (\langle \Psi_{\rm in}^{\rm comp} | \mathbf{A}^{\dag} \mathbf{A} | \Psi_{\rm in}^{\rm comp} \rangle)$, where the maximum and minimum are taken over the set of properly normalized input states. As a more convenient measure, we use the norm $\langle \mathbf A | \mathbf A \rangle$. It is easy to verify that $\{ \mathbf{A}\}_{\rm min}^2 \leq \langle \mathbf A | \mathbf A \rangle \leq \{ \mathbf{A}\}_{\rm max}^2 $. As fidelity $F \rightarrow 1$, $ \{ \mathbf{A}\}_{\rm min} / \{ \mathbf{A}\}_{\rm max} \rightarrow 1$ and $S$ becomes independent of the initial state. We refer to $S = \langle \mathbf A | \mathbf A \rangle$ as {\it the} success probability, keeping in mind that it may not correspond to the success probability for every initial state, except in the case of perfect fidelity.

Once the success rate $S(\mathbf U ) = \langle \mathbf A(\mathbf U ) | \mathbf A(\mathbf U ) \rangle$ and fidelity $F( \mathbf U)$ have been constructed for a given target transformation and given ancilla resources, the task is to find the unitary matrix $\mathbf U$ that maximizes $S(\mathbf U)$ on the constraint set $F(\mathbf U)=1$.  We parametrize $\mathbf U=\exp{(\sum_{j=1}^{N^2}x_j \mathbf{H}_j)}$, where $\mathbf{H}_j$ is a complete set of complex anti-Hermitian $N \times N$ matrices, and find a local maximum of $F$ in $x$ space. When the optimization has converged, we check whether $F(\mathit{ \mathbf U})=1$ to numerical accuracy, in which case we proceed to find a local maximum of success probability $S(\mathbf{U})$ on the $F(\mathbf U)=1$ surface. Repeating the process with multiple randomly chosen starting points, we obtain the best $S$, which yields the optimal design for the quantum circuit.

We first apply our approach to the CS gate. For an arbitrary two-qubit gate, $\mathbf A ( \mathbf U )$ and $\mathbf{A}_{\rm t}$ are $10 \times 4$ matrices;
for the  CS gate matrix $\mathbf{A}_{\rm t}$ is determined according to the action of the gate  $|0101 \rangle  \to - \alpha |0101 \rangle $, and
$|\Psi_{\rm in}^{\rm comp} \rangle \to \alpha|\Psi_{\rm in}^{\rm comp } \rangle$
for $\langle 0101 | \Psi_{\rm in}^{\rm comp } \rangle=0$. The variable $\alpha$ is an arbitrary non-zero complex constant, in agreement with the standard definition of a projective complex space. Here, one easily checks that the minimum number of unentangled ancillas needed to obtain perfect fidelity is $N_a=2$, so that $\mathbf U$ is a $6 \times 6$ matrix ($N=N_c+N_a=4+2=6$). In this case, we find that the second optimization stage is unnecessary, i.e., $S(\mathbf U )$ is a constant on every $F=1$ manifold (each such manifold consisting of an equivalence class of matrices differing only by phase factors). Several inequivalent $F=1$ manifolds are found. The best solutions have $S=2/27$, corresponding to an analytic solution found previously by Knill~\cite{KnillCZ02}. Due to the complexity of the CS gate, it is not known if an analytical proof for determining the maximum success probability is possible. Our numerical evidence, however, strongly indicates that Knill's solution is indeed the global maximum.

Can the solution be improved by adding $N_v$ vacuum modes to the device? This question may be answered straightforwardly by repeating the above optimization with $(N_c + N_a +N _v) \times (N_c + N_a + N_v)$ unitary matrices $\mathbf U$, for various values of $N_v$. However, there exists an alternative ``unitary dilation'' approach~\cite{VanMeter06}.

The most general device design, allowing for an arbitrary number of vacuum modes, is obtained by allowing $\mathbf U$ to be an arbitrary complex $(N_c + N_a) \times (N_c + N_a)$ matrix, and replacing $\mathbf U \to \mathbf U/\| \mathbf U \|$, which scales the maximum singular value to unity. The expression (\ref{Eq:F}) for fidelity as a function of $\mathbf U$  is invariant under scaling, while the generalized success function for a nonunitary $\mathbf U$ is given by
\begin{equation}
\label{singsuccess}
\tilde S(\mathbf U)=S(\mathbf U)/ (\| \mathbf U \|)^{2(M_c+M_a)} \,.
\end{equation}
$\tilde S(\mathbf U)$ has a discontinuous gradient whenever the largest singular value of $\mathbf U$ goes through a double or higher-order degeneracy. Of particular interest is the fact that $\tilde S$, while well-behaved on the manifold of unitary $\mathbf U$, has a singular cusp-like structure in the neighborhood of this manifold.

\begin{figure}[ht]
\includegraphics[width=8.8cm]{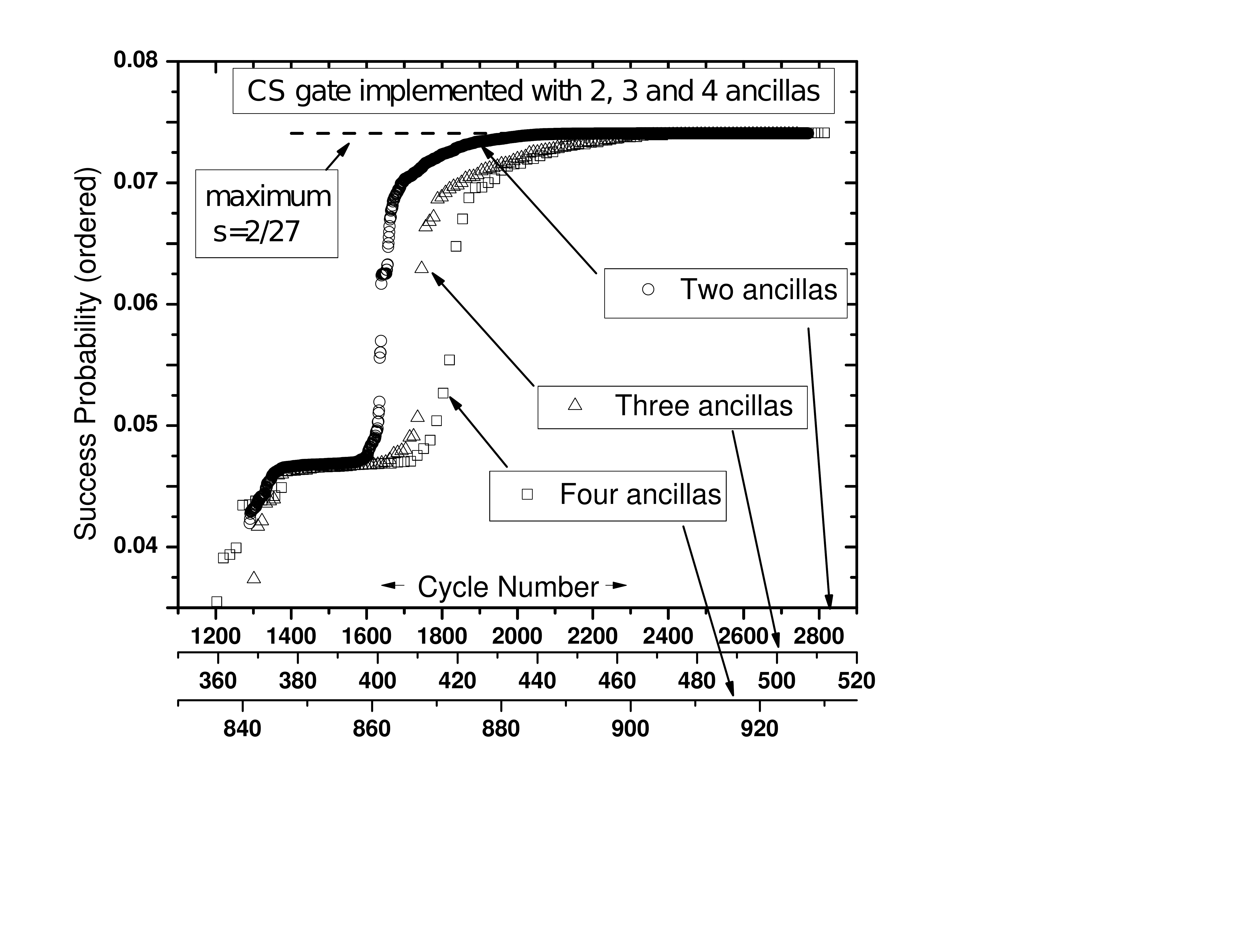}
\caption{\label{fig:CSsuccess} The optimized success probability for the CS gate is shown for two, three, and four ancillas, and an arbitrary number of vacuum modes. Each point indicates a complete run starting from a randomly chosen starting matrix $\mathbf{U}$. The success rates are arranged in ascending order, so that the horizontal axis may be viewed as a cumulative frequency. The $2/27 \approx 0.074$ success rate found by Knill~\cite{KnillCZ02} is indicated by a horizontal line.}
\end{figure}

The result of a non-unitary optimization for the CS gate with two ancillas is shown in Fig.~\ref{fig:CSsuccess} (circles). Plateaus are clearly visible, corresponding to local maxima of the success rate. Indeed, we have continued the optimization for a much larger number of iterations with higher numerical accuracy for data points characterized by $S>0.04$. In doing so, we found that the success rate always converges either to the Knill solution ($S=2/27 \approx 0.074$) or to one of two local maxima: $S \approx 0.047$ and $S =1/16 = 0.0625$. The Knill solution has the largest basin of attraction, while the basin of attraction for $S=1/16$ is the smallest. We note that the KLM scheme~\cite{KLM01} produces a success rate $S=1/16$, though the matrices obtained numerically have a structure different from the KLM form. The pronounced plateau at $2/27$ provides strong numerical evidence that Knill's solution (which makes no use of vacuum modes) is globally optimal, even when vacuum modes are allowed. It appears that the cusp-like structure of the success rate (\ref{singsuccess}) strongly favors maxima appearing at unitary values of $\mathbf U$ (where all singular values become degenerate at $1$), and indeed the global maximum corresponds to one such unitary matrix: the Knill matrix.

Interestingly, analytical fidelity-preserving transformations, constructed by extending the Gr\"obner basis method~\cite{VanMeter06}, can explain seven dimensions of the $F=1$ subspace, while direct numerical tests reveal that this subspace is 11-dimensional in the vicinity of the Knill solution. Thus, there exist hidden symmetries, which, we believe, can be identified only by more powerful mathematical methods from the repertoire of algebraic geometry.

Next, we investigate the effect that additional ancilla resources may have on the optimization problem. Previously, an upper bound for the success probability with unentangled resources was shown to be $3/4$~\cite{Knill03}. Repeating our optimization procedure in larger matrix spaces associated with three and four ancillas (Fig.~\ref{fig:CSsuccess}), we find, surprisingly, that the global and local maxima are unchanged. This suggests that the minimum resources needed to produce the CS gate with perfect fidelity (i.e., two unentangled ancillas with no vacuum modes) also suffice to produce the best possible success rate. In view of the fact that exactly the same behavior of success probability has been found for the NS gate~\cite{EisertNS05}, one may expect that this may be a universal property of probabilistic (photonic) gates: {\it the maximal success probability is attained with minimal required resources}. We tested this conclusion also for the Toffoli gate, which we discuss below in detail, finding that adding one more ancilla to the required minimum of three ancilla photons also does not affect any of two local maxima, in full compliance with the suggested rule.

\begin{figure}[ht]
\includegraphics[width=8.8cm]{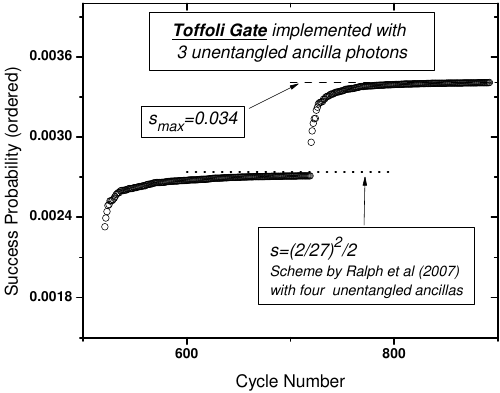}
\caption{\label{fig:Tofsuccess} The distribution of success rates for the Toffoli gate.}
\end{figure}

Now we consider the three-qubit Toffoli gate. After a local Hadamard rotation, the standard Toffoli gate acts as a ``sign" transformation:
$|010101 \rangle \to  \alpha|010101 \rangle $, and
$|\Psi_{\rm in}^{\rm comp} \rangle \to -\alpha|\Psi_{\rm in}^{\rm comp} \rangle$
for $\langle 010101 | \Psi_{\rm in}^{\rm comp} \rangle=0$. We first check that a minimum of three ancillas are needed for perfect fidelity. Thus $N=N_c+N_a=6+3=9$.

To reduce the size of the parameter space, and improve the convergence of the success optimization, we consider the following ansatz
for $\mathbf U$: $U_{ij}=U_{ji}=\delta_{ij}$ for $i=2$, $4$, $6$, i.e., $\mathbf U$ is designed to act non-trivially only on the computational modes 1, 3, and 5.

The results of an optimization over $9 \times 9$ subunitary matrices are shown in Fig.~\ref{fig:Tofsuccess}. The best solution obtained is $S \approx 0.00340$;
This is an improvement over combining a CNOT gate, a CS gate, and a ``passive quantum filter'' to produce the Toffoli gate~\cite{Ralph2007}, which yields a total success rate $S=(2/27)^2 \times {1 /2} \approx 0.00274$ using four unentangled ancilla photons.

Of practical interest is our finding that optimization in the full $10 \times 10$ matrix space is much more efficient than in $13 \times 13$ {\it unitary} space, even though the optimal $N_v=3$ solution is an element of both. The space of unitary $\mathbf{U}$ contains many local maxima of the success rate, preventing the global maximum from being reached. Eliminating the unitarity constraint creates passageways connecting the local maxima to the global one.

In this work, we have provided numerical evidence that the previously obtained solution for the CS (CNOT) gate, with a success probability $S=2/27$, is optimal, and cannot be improved by adding ancillas or auxiliary vacuum modes. On the other hand, for the Toffoli gate we show a new solution, which surpasses what has been obtained analytically using unentangled ancillas (our solution provides a higher success probability using fewer resources). This result is a proof of principle of successful numerical optimization in linear optical quantum information processing. Future directions that naturally suggest themselves include: optimal implementation of two-mode biphotonic qutrit~\cite{Lanyon2008} gates, operations on multi-rail encoded qudits using angular momentum photons~\cite{KwiatAngular2008}, design of gates that are robust to noise and photon loss, and optimization in the context of error-correcting codes~\cite{wildeuskov}.

\begin{acknowledgments}
We are grateful to Mark M. Wilde and Pavel Lougovski for very helpful discussions. This work was supported in part by the NSF under Grants No.\ PHY-0545390 and CRC 0628092, by the Army Research Office, and the Intelligence Advanced Research Projects Activity.
\end{acknowledgments}
	\bibliography{optim_mod2}
\end{document}